\documentclass[
reprint,
 amsmath,amssymb,
 aps, physrev,
]{revtex4-2}

\usepackage{graphicx}
\usepackage{dcolumn}
\usepackage{bm}
\usepackage{xcolor}
\usepackage[left=0.7in,right=0.7in,top=0.6in,bottom=0.65in]{geometry}

\renewcommand{\k}{\mathbf{k}}
\newcommand{\q}{\mathbf{q}}
\newcommand{\p}{\mathbf{p}}
\newcommand{\vel}{\mathbf{v}}

\newcommand{\ketbra}[2]{|#1\rangle\langle #2|}
\newcommand{\braket}[2]{\langle #1 | #2 \rangle}

\newcommand{\bra}[1]{\left<#1\right|}
\newcommand{\ket}[1]{\left|#1\right>}
\newcommand{\OPc}[2]{\hat{#1}_{#2}^{\dag}}
\newcommand{\OP}[2]{\hat{#1}_{#2}^{\vphantom{\dag}}}
\newcommand{\CD}[1]{\OPc{c}{#1}}
\newcommand{\C}[1]{\OP{c}{#1}}

\newcommand{\E}{\epsilon}

\newcommand{\Vol}{\mathcal{V}}
\newcommand{\dq}{\boldsymbol{\delta}}

\DeclareMathOperator{\Real}{\textrm{Re}}
\DeclareMathOperator{\Imag}{\textrm{Im}}

\newcommand{\sectiontitle}[1]{\textit{#1} --}

\newcommand{\supplement}[1]{\cite{supp}} 

\begin{document}

\title{Probing the Quantum Geometry of Correlated Metals using Optical Conductivity} 

\author{Deven P. Carmichael}
\affiliation{Department of Physics and Astronomy, University of Pennsylvania, Philadelphia, PA 19104, USA}
\author{Martin Claassen}%
\affiliation{Department of Physics and Astronomy, University of Pennsylvania, Philadelphia, PA 19104, USA}

\date{\today}

\begin{abstract}
Recent studies have revealed that the quantum geometry of electronic bands determines the electromagnetic properties of non-interacting insulators and semimetals. However, the role of quantum geometry in the optical responses of interacting electron systems remains largely unexplored. Here we examine the interplay between Coulomb interactions and Bloch-band quantum geometry in clean metals. We demonstrate that the low-frequency optical conductivity of a correlated metal encodes the structure of Bloch wavefunctions at the Fermi surface. This response originates from integrating out highly off-resonant interband scattering processes enabled by Coulomb interactions. The resulting quantum-geometric contribution appears generically in multiband systems, but becomes the dominant effect in the optical conductivity for a parabolic band. We consider a dilute correlated metal near a topological band inversion and show that the doping dependence of optical absorption can measure how the orbital character of Bloch wavefunctions changes at the Fermi surface. Our results illustrate how the confluence of quantum geometry and Coulomb interactions can enable optical processes and enrich the physics of Fermi liquids.
\end{abstract}

\maketitle

Quantum-geometric properties of electronic bands have recently emerged as versatile ingredients for understanding and engineering material properties. In band insulators and semimetals, non-linear electromagnetic responses have been shown to depend on the geometric structure of the bands' Bloch wavefunctions, described by the Berry curvature and its multipole moments, the shift vector, the quantum metric, and its generalizations \cite{TKNN_1982,Haldane_Hall,morimoto2016,sodemann2015quantum,ma2021topology,orenstein2021topology,ahn2022riemannian,dejuan2017,Chan_Photocurrents_2017,chaudhary2022,gao2023,komissarov2024quantum, verma2024instantaneous, verma2025framework,Baltz_shift_1981,tan2016shift, du2021nonlinear, Chan_Photocurrents_2017, nagaosa2022nonlinear,morimoto2023geometric,debeule2023berry,torma2023essay, jiang2025revealing}.
Much of this understanding however relies on free-electron descriptions, where interactions are treated at density-functional or mean-field level. In many-body systems, investigations of quantum geometry and electronic correlations instead have primarily focused on ground state properties of fractional Chern insulators \cite{roy2014,claassen2015,jackson2015,leeclaassen2017,wang2021,ledwith2023,liu2024} and flatband superconductors \cite{peotta2015,liang2017,iskin2018,julku2021,torma2022,huhtinen2022,herzogarbeitman2022,chen2024,kitamura2023,Shavit2025,Jahin2026}. Additionally, it is now appreciated that quantum geometry can stabilize flavor ferromagnetism \cite{Wu2020b, Kang2024b, hu2025, chen2026, ding2026}, enhance the Kohn-Luttinger mechanism for superconductivity \cite{Shavit2025, Jahin2026}, and determine symmetry broken phases through quantum geometric nesting \cite{Han2024, Zhang2025}. However, the effect of the structure of Bloch wavefunctions on the {\textit{electrodynamics} of correlated materials remains relatively unexplored. Recent works showed that the Drude weight in flat bands \cite{Ohad_FlatBandDrude}, nonequilibrium control of electronic phases \cite{tai_correlated, gassner2024}, corrections to the Hall conductivity of metals \cite{pasqua2024fermi}, and sum-rules \cite{Souza2000, Onishi2024, Mendez-Valderrama2024, Mao2024, qiu2025quantum} arise from the interplay between quantum geometry and correlations. However, whether Bloch-state quantum geometry in correlated electron systems can affect and be probed by terahertz electromagnetic responses beyond sum rules remains an intriguing open question. 

In this Letter, we address this question by studying the optical conductivity $\sigma(\omega)$ of correlated metals. We find a new contribution to the low-frequency scaling behavior of $\Real \sigma(\omega)$ which originates from the quantum geometry of Bloch states at the Fermi surface and is mediated via electron-electron scattering. In systems with non-trivial quantum geometry, such as near a topological band inversion, this leads to a dramatic enhancement of the low-frequency optical conductivity. At dilute filling close to the bottom of a band, we show that $\Real \sigma(\omega)$ has a purely quantum-geometric origin by virtue of a nearly parabolic dispersion which we illustrate for a class of interacting metals near a higher-order topological band inversion. Our results demonstrate how optical responses can originate entirely from the interplay between quantum geometry and interactions away from flat-band limits. Additionally, these findings offer a new route to experimentally probe the quantum geometry of interacting electrons via THz optical conductivity.

\begin{figure}
\includegraphics[width=.95\columnwidth,trim=0cm 0.1cm 0cm 0.0cm,clip]{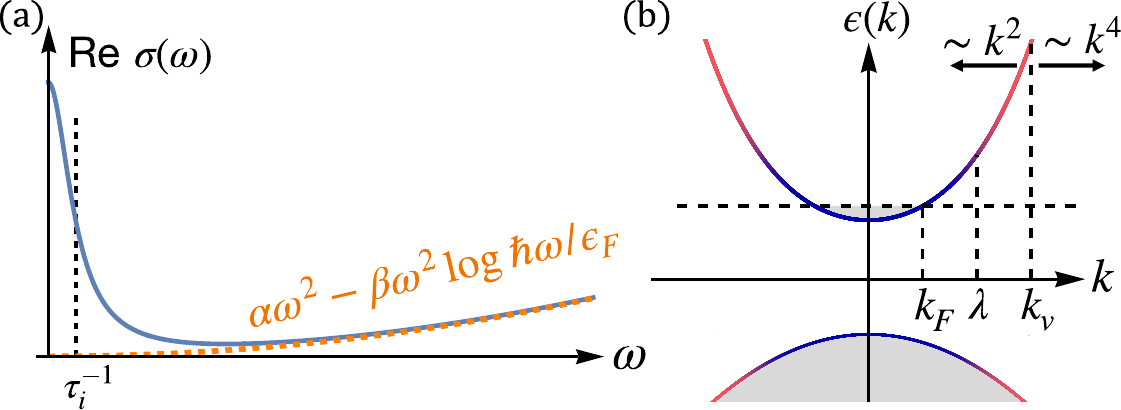}
\caption{\label{fig:overview} (a) Schematic of the optical conductivity of a correlated metal. At low frequencies the conductivity is dominated by a Drude peak, while in the hydrodynamic regime the conductivity scales quadratically (with a possible logarithmic correction) for an isotropic 2D metal. (b) Momentum scales that govern dilute metals near a band inversion: Fermi wavevector $k_F$, band inversion scale $\lambda$ where the orbital character changes, and wavevector $k_v$ for nonparabolic corrections.}
\end{figure}

\sectiontitle{Optical conductivity of clean metals} In a metal, $\Real \sigma(\omega)$ probes distinct current relaxation mechanisms at different frequencies [Fig.~\ref{fig:overview}(a)], which can be used to disentangle individual scattering processes. In clean systems, a narrow Drude peak at $\omega = 0$ has a width set by impurity scattering $\tau_i^{-1}$ and temperature $T$. For frequencies $\omega > \tau_i^{-1}, T$ but below any interband transitions or collective modes, the response lies in the hydrodynamic regime and is governed by electron-electron scattering \cite{Maslov_OpticalCond}. Here, $\Real \sigma(\omega)$ is generically given by the Gurzhi formula \cite{gurzhi1959,Maslov_OpticalCond} which describes a constant background at finite frequencies. However deviations from Gurzhi scaling can occur for special Fermi surface geometries. For example, in the absence of Umklapp scattering, the optical conductivity of a small convex Fermi surface in two dimensions scales as $\omega^2$ with a possible $\omega^2 \log \omega$ term~\cite{rosch2005zero,rosch2006optical,Sharma_DFL, mu2024adequacy}, as shown in Fig.~1(a). 

The most drastic departure from Gurzhi scaling occurs in Galilean invariant metals where the conductivity vanishes identically \cite{Maslov_OpticalCond, mu2022optical}. This is typically expected to occur for all parabolic dispersions, such as near the top or bottom of a band at dilute filling [Fig.~\ref{fig:overview}(b)]. Here, conservation of momentum enforces that Coulomb scattering pairs of electrons $(\k,\k') \to (\k+\q,\k'-\q)$ cannot change the net electron velocity $\vel_{\k+\q} + \vel_{\k'-\q} - \vel_{\k'} - \vel_{\k} = 0$. Hence, current-carrying excitations cannot be generated from the ground state and the optical conductivity is zero. For dilute metals, the optical conductivity will therefore only become finite once the Fermi vector $k_F$ is doped past the scale $k_v$ at which the band is no longer parabolic [Fig.~\ref{fig:overview}(b)]. However, we will show there is a loophole in this argument: the \textit{change} of the Bloch wavefunctions at the Fermi surface breaks Galilean invariance, even though the dispersion does not, enabling a finite current in the presence of interactions. Instead of vanishing, the optical response of dilute metals becomes strongly enhanced when the Fermi surface lies at a new momentum scale $\lambda$ of maximal orbital character change [Fig.~\ref{fig:overview}(b)] even if the band is parabolic. This hitherto missed and purely quantum geometric contribution to the conductivity manifests in a new filling-dependent maximum in low-frequency absorption, suggesting that THz spectroscopy of metals as a function of a gate-controlled Fermi level can interrogate the quantum geometry of correlated electron systems.

\begin{figure}
\includegraphics[width=\columnwidth, trim=0cm 0.1cm 0cm 0.0cm,clip]{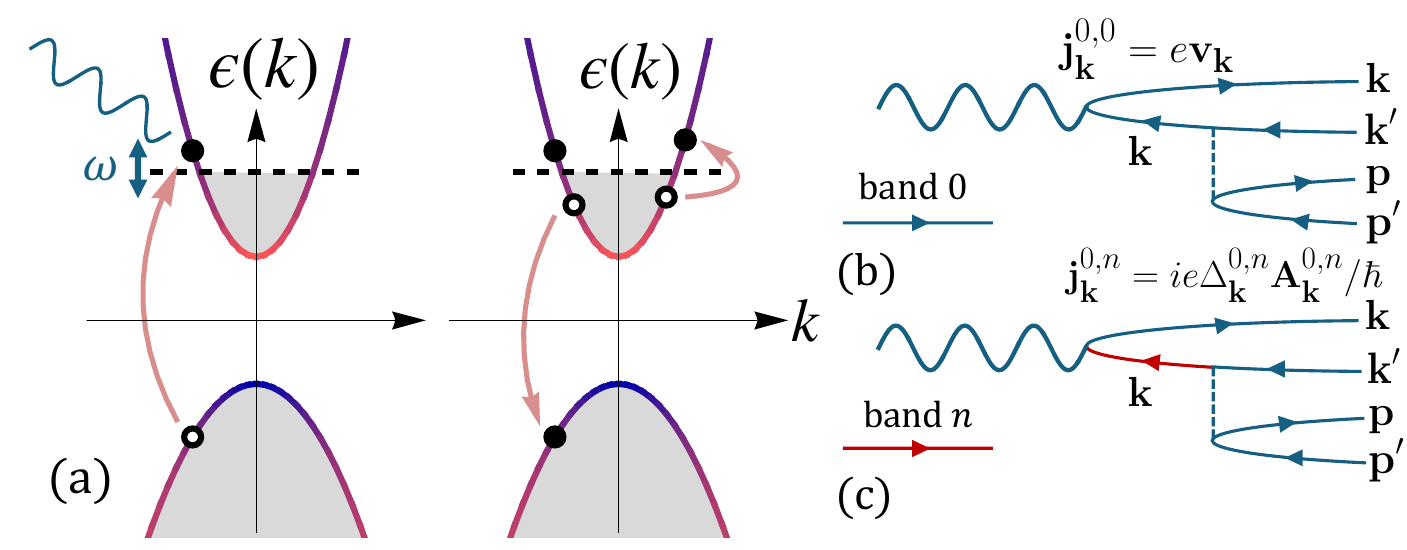}
\label{fig:diagrams}
\caption{\label{fig:diagram} (a) Schematic of an optical absorption process which results in a quantum geometric contribution to $\Real \sigma(\omega)$. (b) Diagram for an intraband excitation from a velocity-dependent intraband current vertex $\mathbf{j}^{0,0}_\k \sim \vel_\k$. (c) Quantum-geometric diagram with an \textit{inter}band-excited intermediate state, corresponding to the processes depicted in (a). The interband current $\mathbf{j}^{0,n}_k$ is proportional to the excitation gap $\Delta^{0,n}_\k$, and the interband Berry connection $\mathbf{A}^{0,n}_\k$.}
\end{figure}

\sectiontitle{Model} To illustrate this effect, consider a two-dimensional metal with screened Coulomb interactions $U_\q$. For simplicity, we assume inversion and time-reversal symmetry. We take the bands to be either spinless or spinful, leading to spin degenerate bands. The material then has bands with dispersion $\E_{n,\k}$ and Bloch wavefunctions with cell-periodic components $u_{\k,\sigma}^n(\mathbf{r})$, where $n$ and $\sigma$ are band and spin indices, respectively. A small Fermi surface is hosted by a single band with dispersion $\E_{0,\k}$ as depicted in Fig.~\ref{fig:overview}(b). All other bands are fully filled or entirely empty. The Hamiltonian for such a correlated metal is given by
\begin{align}
    \hat{H} = &\sum_{\k n \sigma}
    \epsilon_{n,\k}~
    \CD{\k,n,\sigma} \C{\k,n,\sigma}
    + 
    \frac{1}{2 \Vol} \sum_{\q} U_\q :
    \rho_{\q} \rho_{-\q}
    :
\end{align}
where $:~:$ denotes normal ordering and
$\rho_\q =
\sum_{\k m n \sigma}
\langle u_{\k+\q, \sigma}^{n} | u_{\k, \sigma}^{m}  \rangle
\CD{\k+\q,n,\sigma}
\C{\k,m,\sigma}$
are the density operators, and $\mathcal{V}$ is the system size.

\sectiontitle{Optical Conductivity} We now compute the real part of the low-frequency longitudinal optical conductivity $\sigma_{\mu\mu}(\omega)$. The paramagnetic current operator
$\hat{\mathbf{J}} = \sum_{\k n m \sigma} \mathbf{j}^{n,m}_{\k \sigma} \CD{\k,n,\sigma} \C{\k,m,\sigma}$ can be decomposed into \textit{intra}- and \textit{inter}-band contributions\cite{blount1962}\footnote{A quick way to derive this is to note that the current operator is related to the momentum derivative of the Bloch Hamiltonian by $\mathbf{j}_{\mathbf{k}\sigma} = \frac{e}{\hbar} \boldsymbol{\nabla}_{\k}h_{\k \sigma} $. With this we can rewrite 
$
\bra{u^n_{\k, \sigma}}  \boldsymbol{\nabla_{\k}}h_{\k \sigma} \ket{u^m_{\k, \sigma}}
=  \boldsymbol{\nabla_{\k}}
\bra{u^n_{\k, \sigma}}  h_{\k \sigma} \ket{u^m_{\k, \sigma}}
- \bra{\boldsymbol{\nabla_{\k}}u^n_{\k, \sigma}}  h_{\k \sigma} \ket{u^m_{\k, \sigma}}
- \bra{u^n_{\k, \sigma}}  h_{\k \sigma} \ket{\boldsymbol{\nabla_{\k}}u^m_{\k, \sigma}} 
= \hbar v^{n}_\k \delta_{n,m} + i \Delta_{\k}^{n,m}
    \mathbf{A}_\mathbf{k\sigma}^{n,m}
$
to arrive at Eq. (2) in the main text.
}
\begin{align}
    \mathbf{j}_{\mathbf{k}\sigma}^{n,m} = e \mathbf{v}_{\mathbf{k}}^{n} \delta_{n,m} + i\frac{e}{\hbar}
    \Delta_{\k}^{n,m}
    \mathbf{A}_\mathbf{k\sigma}^{n,m}
\end{align}
where $e$ is the electron charge, $\mathbf{v}_\mathbf{k}^{n} = \frac{1}{\hbar}\boldsymbol{\nabla_{\k}} \epsilon_{n,\mathbf{k}}$ is the velocity, $\Delta_{\k}^{n,m} = \epsilon_{n,\mathbf{k}} - \epsilon_{m, \mathbf{k}} $ are band gaps at $\k$, and $\mathbf{A}_\mathbf{\k\sigma}^{n,m} = i \left \langle u_{\mathbf{k}\sigma}^n | \boldsymbol{\nabla_{\k}} u_{\mathbf{k}\sigma}^m \right \rangle $ are interband Berry connections.

At small frequencies and low temperatures $k_B T \ll \hbar\omega \ll \Delta$ (where $\Delta$ is the lowest energy for interband transitions from the Fermi energy), $\Real \sigma_{\mu\mu}$ will probe electron-electron scattering at the Fermi surface. At first glance, this suggests that processes which involve off-resonant interband transitions would not contribute provided that $\omega$ is much smaller than the transition energy. However this is not the case. Consider the process shown in Fig.~\ref{fig:diagram}(a), where a photon excites a virtual interband excitation to a remote band $n$, followed by Coulomb scattering, resulting in a final state of two particle-hole excitations near the Fermi surface and resonant with the incident photon.
Despite the highly off resonant intermediate state, such processes scale e.g. as
$\propto \frac{\Delta^{0,n}_\k}{\Delta^{0,n}_\k - \hbar\omega} \mathbf{A}_{\mathbf{k}}^{0,n} U_\q \braket{u_{\k,\sigma}^n}{u_{\k-\q,\sigma}^0} \approx \mathbf{A}_{\mathbf{k}}^{0,n} U_\q \braket{u_{\k,\sigma}^n}{u_{\k-\q,\sigma}^0} + \mathcal{O}(\hbar \omega / \Delta)$
. Summed over all interband transitions and using the identity $\sum_{n\neq 0}\ketbra{u_{\k,\sigma}^n}{u_{\k,\sigma}^n} = 1 - \ketbra{u_{\k,\sigma}^0}{u_{\k,\sigma}^0}$, these processes can be expressed solely as properties of the partially occupied band and must enter as quantum-geometric contributions into the low-energy (adiabatic) theory of correlated metals.

To make this argument quantitative, we compute $\Real \sigma(\omega)$ using Fermi's golden rule (FGR) to leading order in Coulomb scattering $U_\q^2$. FGR has been used to study the spatially dispersive optical conductivity in graphene \cite{Mishchenko_Plasmon, Sharma_Plasmon} as well as $ \Real \sigma(\omega)$ due to intervalley scattering \cite{Gindikin_Intervalley}.
The final state is reached by two families of diagrams, those with intraband [Fig.~2(b)] or interband [Fig.~2(c)] excitations as intermediate states. 
We find
\begin{widetext}
\begin{align}
\label{eq:fullresponse}
\Real \sigma_{\mu\mu} \left( \omega \right) =
\frac{\pi e^2}{4 \omega \Vol^3} \left( 1 - e^{-\beta \omega} \right ) 
 &
\sum_{\k\p\q} \sum_{\sigma\sigma'} \left| \left[
\boldsymbol{\mathcal{M}}_{\mathbf{k},\mathbf{p},\mathbf{q}}^{\sigma\sigma'}
-
\delta_{\sigma, \sigma'}
\boldsymbol{\mathcal{M}}_{\k,\p,\p-\k-\q}^{\sigma\sigma'}
\right]_\mu \right|^2 
\nonumber
\\
& 
\times~ 
n(\epsilon_{\k})
n(\epsilon_{\mathbf{p}} )
\left [ 1 - n(\epsilon_{\mathbf{p}-\q}) \right ]
\left [ 1 - n(\epsilon_{\k+\q}) \right ]
\delta \left (
\epsilon_{\k+\q} + \epsilon_{\mathbf{p}-\q}
 - \epsilon_{\mathbf{p}} - \epsilon_{\k} -\hbar\omega
\right )
\end{align}
\end{widetext}
with a matrix element  \supplement{Supplementary Material, Section A} that has two contributions $\boldsymbol{\mathcal{M}}_{\mathbf{k}, \mathbf{p}, \mathbf{q}}^{\sigma\sigma'} 
=
\boldsymbol{\mathcal{M}}_{\mathbf{k}, \mathbf{p}, \mathbf{q}}^{\sigma\sigma'; \textrm{vel}}
+ 
\boldsymbol{\mathcal{M}}_{\mathbf{k}, \mathbf{p}, \mathbf{q}}^{\sigma\sigma'; \textrm{geo}}
$
, one from conventional intraband absorption and one from quantum geometry:
\begin{align}
\label{eq:M_vel}
\boldsymbol{\mathcal{M}}_{\mathbf{k}, \mathbf{p}, \mathbf{q}}^{\sigma\sigma'; \textrm{vel}}
=~&
- \frac{U_\mathbf{q}}{\hbar \omega} 
\left (
\mathbf{v}_{\mathbf{k}+\mathbf{q}}
+ \mathbf{v}_{\mathbf{p}-\mathbf{q}} 
- \mathbf{v}_{\mathbf{k}}
- \mathbf{v}_{\mathbf{p}}
\right )  \notag\\
&\times 
\langle u_{\mathbf{k}+\mathbf{q}, \sigma} | u_{\mathbf{k}, \sigma} \rangle 
\langle u_{\mathbf{p}-\mathbf{q},\sigma'} | u_{\mathbf{p}, \sigma'} \rangle
\\
\label{eq:M_geo}
\boldsymbol{\mathcal{M}}_{\mathbf{k}, \mathbf{p}, \mathbf{q}}^{\sigma\sigma'; \textrm{geo}} =~&
\frac{1}{\hbar} U_\mathbf{q} \ \boldsymbol{\mathcal{D}}
\langle u_{\mathbf{k}+\mathbf{q}, \sigma} | u_{\mathbf{k}, \sigma} \rangle 
\langle u_{\mathbf{p}-\mathbf{q},\sigma'} | u_{\mathbf{p}, \sigma'} \rangle ~.
\end{align}
Here, $\boldsymbol{\mathcal{D}}$ denotes a \textit{generalized} covariant derivative whose action on Bloch states reads $\boldsymbol{\mathcal{D}} | u_{\k, \sigma} \rangle = 
( 
\boldsymbol{\nabla}_{\k} + i \mathbf{A}_{\k,\sigma}
) | u_{\k, \sigma} \rangle
$ and $\boldsymbol{\mathcal{D}} \langle u_{\k,\sigma} | = 
\left ( 
\boldsymbol{\nabla}_{\k} - i \mathbf{A}_{\k,\sigma}
\right ) \langle u_{\k, \sigma} | $, where $\mathbf{A}_{\k, \sigma} =i \langle u_{\k, \sigma} | \boldsymbol{\nabla_{\k}} u_{\k, \sigma} \rangle $ is the \textit{intra}band Berry connection for the active band and we have dropped the band index as all expressions only depend on the active band.

Eq.~(\ref{eq:M_vel}) describes conventional intraband absorption $\sim e \mathbf{v}_\k$ [Fig.~\ref{fig:diagram}(b)] and recovers prior results for $\Real \sigma(\omega)$ of correlated metals \cite{Sharma_DFL, Gindikin_Intervalley}, with the Coulomb interaction picking up a form factor from the Bloch wavefunctions. Importantly, however, Eq. (\ref{eq:M_vel}) vanishes for a parabolic band. Meanwhile, the quantum geometric contribution [Eq.~(\ref{eq:M_geo})] is new and remains finite. It arises from carefully integrating out virtual interband excitations [Fig.~\ref{fig:diagram}(c)].
At dilute filling near a parabolic band bottom, the optical conductivity of a metal therefore becomes solely a function of the \textit{change} of the Bloch states at the Fermi surface, parameterized by their quantum geometry.

\sectiontitle{Parabolic Limit} To illustrate this, consider first a perfectly parabolic band. For a convex two-dimensional Fermi surface, the low-frequency optical response is governed by exchange channel ($\k, \p + \dq \mapsto \p, \k+\dq$) and pairing channel ($\k, -\k + \dq \mapsto \p, -\p + \dq$) scattering, where $\dq$ is a small momentum shift away from the Fermi surface [see insets of Fig.~\ref{fig:geocon}]. The low-frequency scaling of the optical conductivity can be computed analytically by taking $\dq \to 0$ \supplement{Supplementary Material, Section B}.
For $\omega \gg k_B T$, we find that the optical conductivity is given by a Fermi surface integral
\begin{align}
    \Real \sigma_{\mu\mu}(\omega) = \frac{e^2}{h^3} \omega^2 ~\frac{1}{48 (2\pi)^2} \oint\limits_{\textrm{FS}^*} \frac{d\k d\p ~ U^2_{\k-\p} }{ v_\k v_\p
\left |\vel_\k \times \vel_\p \right | } ~\mathcal{G}_{\k,\p}^{\mu}  \label{eq:sigma_FS}
\end{align}
where $\textrm{FS}^*$ excludes possible divergent contributions discussed below, and $\mathcal{G}_{\k,\p}^{\mu}$ is a two-particle Fermi surface quantum-geometric tensor
\begin{align}
    &\mathcal{G}_{\k,\p}^{\mu} = \sum_{\sigma} \left\{ \left| (\partial_{k_\mu} + \partial_{p_\mu}) \left| \langle u_{\p, \sigma}| u_{\k, \sigma} \rangle \right|^2 \right|^2 \right. \notag\\
        &+ \left| \left[ \partial_{k_\mu} + \partial_{p_\mu}
        - 2 i ( A^\mu_{\p, \sigma} - A^\mu_{\k, \sigma} )
        \right]
    \langle u_{\p,\sigma}| 
    u_{\k,\sigma} \rangle
    \langle u_{\k,-\sigma}| 
    u_{\p,-\sigma} \rangle 
    \right 
    |^2 \notag\\
    & + 
    | 2i \Imag \left [
    \langle u_{\k,\sigma}| 
    u_{\p,\sigma} \rangle
    ( \partial_{k_\mu} + \partial_{p_\mu})
    \langle u_{\p,\sigma}| 
    u_{\k,\sigma} \rangle
    \right]
    \notag \\
    & \left.\mspace{20 mu}
    - 2i ( A^\mu_{\p,\sigma} - A^\mu_{\k, \sigma} )
    \langle u_{\k,\sigma}| 
    u_{\p,\sigma} \rangle \langle u_{\p,\sigma}| 
    u_{\k,\sigma} \rangle
    |^2 \right\} \label{eq:channels}
\end{align}
Here, the three terms describe same-spin exchange, opposite-spin exchange and opposite-spin pairing channel scattering, respectively. The same-spin pairing-channel $\dq \to 0$ matrix element vanishes under inversion symmetry. We assume inversion and time-reversal symmetry, so that $| u_{-\k, -\sigma} \rangle = | u_{\k, \sigma}^* \rangle, \ \mathbf{A}_{-\k,\sigma} = -\mathbf{A}_{\k,\sigma}$ and $\mathbf{A}_{-\k,-\sigma} = \mathbf{A}_{\k,\sigma}$, in order to simplify the expressions. For small momentum transfer $\mathcal{G}_{\k,\p}^{\mu}$ is proportional to the Berry curvature for spinful electrons and the derivative of the quantum metric for spinless electrons \supplement{}.

The velocity denominator in Eq.~(\ref{eq:sigma_FS}) effectively measures the phase space for scattering processes \supplement{Supplementary Material, Section B} and can diverge at Fermi surface momenta $\k$, $\p$ where $\vel_\p \parallel \vel_\k$. Physically, these points allow electrons to resonantly scatter tangentially at leading order in $\dq$.  In reality, the curvature of the Fermi surface constrains this scattering. Formally,  the Fermi surface $\rm{FS}^*$ integral of Eq. (\ref{eq:sigma_FS}) must be chosen to exclude these points by imposing a cutoff, proportional to the energy transfer $\omega$ of the photon-assisted scattering process. We write this cutoff as $\omega/\Lambda$ where $\Lambda$ is a constant that solely depends on the dispersion at the Fermi surface. This choice of regularization cleanly separates the optical conductivity
\begin{align}
    \Real \sigma_{\mu\mu}(\omega) = \alpha_\mu ~ \omega^2 - \beta_\mu ~ \omega^2 \log\left(\hbar\omega/\E_F \right)
\end{align}
into a regular ($\alpha$) contribution and a logarithmic ($\beta$) contribution that is determined entirely from the matrix elements for scattering processes where $\vel_\p \parallel \vel_\k$ \supplement{Supplementary Material, Section C}. For a circular Fermi surface, and noting that $\mathcal{G}_{\k,\k}^\mu = 0$, the latter only happens for $\k \leftrightarrow -\k$ scattering which has a matrix element that can be expressed simply in terms of the overlaps of the Bloch states at opposite momenta times the Berry connection plus a term which renders it gauge invariant:
\begin{align}
    \beta_\mu = 
    \frac{e^2}{h^3} & \frac{k_F}{12 (2\pi)^2}  \sum_\sigma\oint\limits_{\text{FS}} \frac{d\k  ~ U^2_{2 \k_F} }{ (v_\k v_{-\k})^2
}
\left | 
\braket{u_{-\k \sigma}}{u_{\k \sigma}}
\right |^4 
\notag 
\\
& 
\times
\left |
2  A^\mu_{\k \sigma} + \left (
\partial_{\k^\mu} + \partial_{-\k^\mu} 
\right)
\textrm{Arg} \braket{
u_{-\k \sigma}
}{
u_{\k \sigma}
}
\right |^2 
\label{eq:log}
\end{align}
As such, quantum geometry restores the $\omega^2$ scaling of the conductivity found for generic small 2D Fermi surfaces \cite{rosch2005zero} while also potentially contributing a term that scales as $\omega^2 \log (\hbar \omega / \E_F)$ as in a Dirac Fermi liquid~\cite{Sharma_DFL}.

While more familiar single-particle quantum geometric objects such as the  Berry connection or quantum geometric tensor depend on only a single momentum, many-body systems must track the scattering of \textit{pairs} of electrons between different momenta. The gauge-invariant two-particle geometric tensor of Eq. (\ref{eq:channels}) therefore reflects the momentum-dependent change of the two-particle scattering states at the Fermi surface. In a perfectly parabolic band these are the only contributions to the conductivity. However, as the Fermi level is moved away from a band minimum, higher order corrections to the Fermi surface dispersion will give an additional contribution from the imperfect cancellation of the velocities in Eq.~(\ref{eq:M_vel}). Due to inversion symmetry, the next leading order correction to the dispersion will go as $\sim \k^4$. For a single Fermi surface, the resulting $\mathcal{M}^{\textrm{vel}}$ contribution keeps the same scaling as the geometric contribution.

\sectiontitle{Dilute Metals near a Band Inversion} To understand the interplay between the contributions from non-parabolicity and quantum geometry for realistic bands, we study a dilute metal near a higher-angular-momentum topological band inversion [Fig.~1(b)]. Without loss of generality, we consider the chiral limit for Bloch Hamiltonians with two orbitals that differ by angular momentum $m$ \cite{Venderbos_2018}:
\begin{equation}
h_\sigma(\k) = \left[\begin{array}{rr}
\E_\k - \Delta (\frac{k}{\lambda})^{2m}  & \Delta (\frac{\sigma k_x + i k_y}{\lambda})^m \\[4 pt]
\Delta (\frac{\sigma k_x - i k_y}{\lambda})^m &
\E_\k - \Delta
\end{array}\right] \label{eq:chiral-inversion}
\end{equation}
Here, $\Delta$ is the band gap, and $\lambda$ is a momentum that parameterizes where the band inversion takes place. The conduction band wavefunction reads
\begin{equation}
    | u_\k \rangle = \frac{1}{\sqrt{(k/\lambda)^{2m} + 1} }
    \left(
    \begin{array}{c}
    1\\
    \left (
    (k_x - i k_y)/\lambda \right )^m
    \end{array}\right)\;
\end{equation}
Furthermore, $\E_\k$ is the conduction band dispersion, which we take to be
\begin{equation}
    \epsilon_\k = \frac{\E_F}{1 + (k_F/k_v)^{4}} \left ( \left( k/k_v \right)^4 + (k/k_F)^2 \right ) - \E_F
\end{equation}
where $k_v$ is the scale above which the band is no longer parabolic [Fig.~1(b)]. $\lambda / k_F$ and $k_v / k_F$ therefore \textit{independently} tune the Bloch wavefunction and dispersion at the Fermi surface.
This model is minimal yet a complete description of the relevant momentum scales of a small Fermi surface and hence applies to a broad class of centrosymmetric materials. Moreover, the model with $m=1$ is equivalent to a massive Dirac fermion, which describes the valleys of transition metal dichalcogenides \cite{PhysRevLett.108.196802}. Here, due to broken inversion symmetry, trigonal distortion $\sim k^3$ leads to breaking of Galilean invariance as well, but can also be treated using our techniques. 

We calculate the optical response of this model by evaluating Eq.~(\ref{eq:fullresponse}) numerically at $T=0$ using Monte Carlo integration \supplement{Supplementary Material, Section C} for a Hubbard interaction $U_{\q}=U$. We then extract the low-frequency scaling by fitting to the functional form $\text{Re} \ \sigma_{\mu\mu}(\omega) = \alpha \omega^2 - \beta \omega^2 \log \left ( \hbar\omega/\E_F \right )$. Due to rotation symmetry, the longitudinal conductivity does not depend on $\mu$ and we drop the index going forward.

We begin by assessing the almost-Galilean-invariant limit, $k_v \gg k_F$, where the Fermi surface is parabolic~\footnote{In the numerics, we take $k_v/k_F = 50$.}. Here, the conventional intraband response is suppressed and $\Real \sigma(\omega)$ is entirely due to quantum geometric scattering. Fig.~\ref{fig:geocon} compares the numerically-determined scaling factors to the analytically-derived quantum-geometric response [Eq.~(\ref{eq:sigma_FS}) and (\ref{eq:log})]. The cutoff $\Lambda$ is in principle derivable from the band dispersion. The value of the cut off contributes to the coefficient of the quadratic scaling form we extract from the numerics because $\beta \omega^2 \log \omega/\Lambda = \beta\omega^2 \log(\E_F/\hbar\Lambda) + \beta \omega^2 \log (\hbar\omega/\E_F)$. As such we fix the value of $\Lambda$ once; the analytical expressions and numerics then match for arbitrary $\lambda/k_F$ and $m$ as the dispersion does not depend on these parameters. In each case we find a good agreement between numerics and analytics.

\begin{figure}[t!]
\includegraphics[width=\columnwidth,trim=0cm 0.5cm 0cm 0cm,clip]{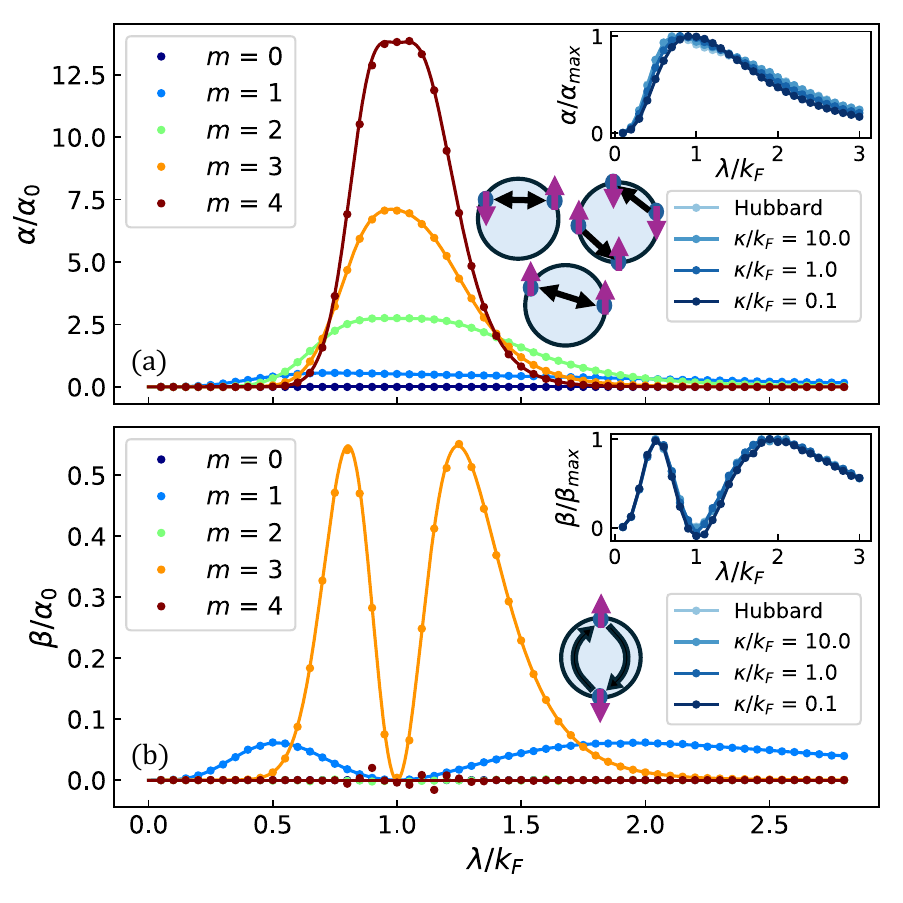}
\caption{\label{fig:geocon} Optical conductivity of dilute correlated metals near a band inversion with angular momentum $m$. Dots are scaling coefficients extracted from the numerically-computed optical conductivity. Lines depict analytical quantum-geometric Fermi surface contributions, where $\alpha_0 = \frac{1}{(2 \pi)^6} \hbar e^2 U^2 k_F^4/\E_F^4$ is an overall scale. The insets are schematics of the relevant scattering channels. (a) Coefficient of $\omega^2$ scaling for spinful fermions where both the exchange and pairing channels contribute. (b) Coefficient of $\omega^2 \log(\hbar\omega/\E_F)$ scaling for spinful fermions, arising from opposite spin scattering between $\k$ and $-\k$. Insets: dependence of scaling coefficients for a $m=1$ band inversion on the Thomas-Fermi screening wavevector $\kappa$.
}
\end{figure}

\begin{figure*}[t!]
\includegraphics[width=2.\columnwidth,trim=0cm 0.4cm 0cm 0.4cm,clip]{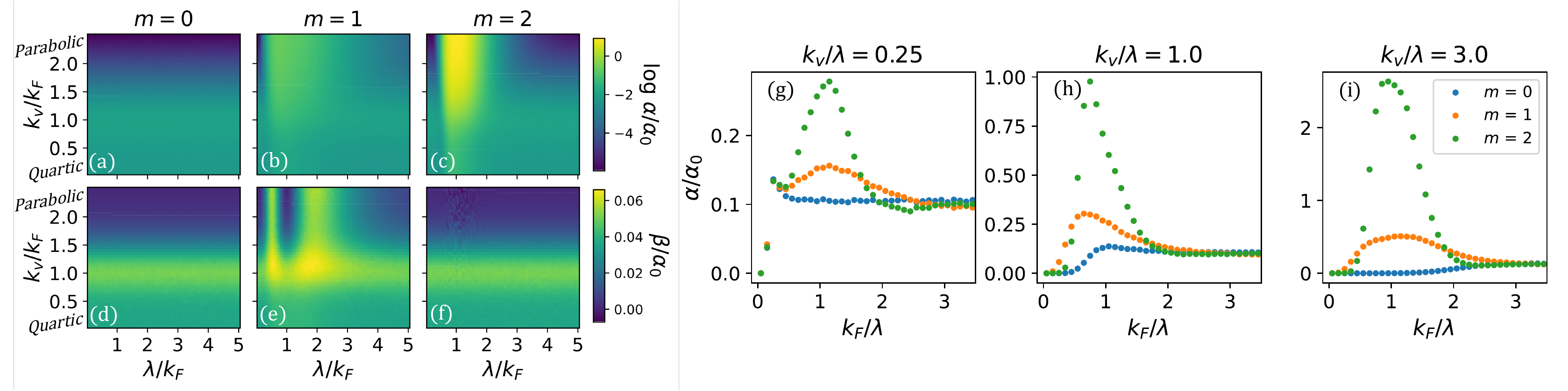}
\caption{\label{fig:fullCond} Scaling contributions for the optical conductivity from non-parabolicity ($k_v$) and quantum geometry ($\lambda$) for spinful fermions, for an isolated band [$m=0$; (a), (d)] and for Fermi surfaces near an angular-momentum $m=1,2$ band inversion [(b),(c),(e),(f)]. (a-c) and (d-f) parameterize $\omega^2$ and $\omega^2 \log(\hbar\omega/\epsilon_F)$ scaling, respectively. In materials, the ratio $k_v/\lambda$ is fixed; however, optical absorption exhibits a maximum as a function of doping [(g-i)], a signature of quantum-geometric correlated metals.}
\end{figure*}

$\alpha$ [Fig.~\ref{fig:geocon}(a)] peaks around $k_F = \lambda$ where the band inversion occurs and increases with higher angular momentum $m$. This reflects that quantum-geometry-enabled optical conductivity is largest when the orbital character of the Bloch wavefunction is maximally changing along the circumference of the Fermi surface.
This can also be understood via the orbital pseudospin texture that describes the orbital character of Bloch states at different momenta near the band inversion. For $\lambda \sim k_F$, the orbital pseudospins at $k_F$ lie along the equator of the Bloch sphere (denoting mixed orbital character) and wind $m$ times as one moves around the Fermi surface, thereby changing maximally along the Fermi surface. Interestingly, the logarithmic term [Fig.~\ref{fig:geocon}(b)] vanishes in the case of even $m$. This can be understood as follows. Since the log term comes entirely from scattering from $\k$ to $-\k$ it reflects how the Bloch wavefunction differs at $\k$ and $-\k$. When $m$ is even, the wavefunctions at $\k$ and $-\k$ are identical (as the band inversion takes place between same-parity orbitals) and the overlap is trivial which means the covariant derivative in Eq.~(\ref{eq:M_geo}) vanishes. Relatedly, the logarithmic term vanishes at $k_F = \lambda$ for odd $m$ as the Bloch functions at $\k$ and $-\k$ are orthogonal. Since the optical conductivity arises from the breaking of Galilean invariance by Fermi surface Bloch functions, it is insensitive to the range of interactions. To demonstrate this, we study the $\sigma(\omega)$ scaling for a $m=1$ band inversion with a Thomas-Fermi interaction $U_\q = U_0/(q+\kappa)$ for various values of the screening wavevector $\kappa$ in the insets of Fig.~\ref{fig:geocon}. Since the magnitude of the response scales with $\kappa$, we rescale $\alpha$ and $\beta$ with respect to their maximum values $\alpha_{\rm max}, \beta_{\rm max}$ to focus on their dependence on $k_F$. We find that both weakly screened (small $\kappa$) and highly screened (large $\kappa$) dilute metals collapse onto the same quantum-geometry-driven doping dependence.

The interplay of Fermi surface quantum geometry ($\lambda/k_F$) and non-parabolicity ($k_v/k_F$) is depicted in Fig.~\ref{fig:fullCond}, as a function of $m$. The $m=0$ case [Fig.~\ref{fig:fullCond}(a),(d)] describes a geometrically-trivial band. Here, results are consistent with previous work \cite{Sharma_DFL}. The conductivity vanishes in the parabolic limit (large $k_v/k_F$) but becomes finite as the band deviates from Galilean invariance at small $k_v/k_F$, and does not depend on $\lambda$.

In contrast, band inversions with nonzero $m$ [Fig.~\ref{fig:fullCond}(b),(c),(e),(f)] enable a finite optical conductivity even when the band is nearly parabolic. Unlike the trivial $m=0$ case, the optical conductivity now exhibits a pronounced enhancement for $\lambda \sim k_F$ when the Bloch wavefunctions maximally change at the Fermi surface. The purely quantum geometric contribution, seen for a parabolic band  ($\sim \k^2$ for $k_F < k_v$), is complemented by a conventional Fermi surface velocity contribution as the dispersion at $k_F$ changes to quartic ($\sim \k^4$ for $k_F > k_v$). For $k_F > k_v$, the optical conductivity decreases due to the reduction of density of states at the Fermi surface, yet the absorption remains peaked near $k_F \sim \lambda$ where the quantum-geometric contribution is maximal.

In a band inversion in a real material, $k_v$ and $\lambda$ are fixed properties of the band structure near the band bottom. Changing the Fermi level via gating traces a diagonal path in Fig.~\ref{fig:fullCond}(a)-(f). The doping dependence near the band inversion scale is shown in Fig.~\ref{fig:fullCond}(g,h,i). Generically, both geometric and dispersive corrections will contribute to the optical conductivity. However, these contributions are distinguished by their dependence on the Fermi level. As the Fermi level is increased from the band bottom, the contribution from  nonparabolicity will gradually turn on and remain relatively constant. Meanwhile, the quantum geometric contribution will peak when the orbital character of the Fermi surface is maximally changing, a telltale signature of quantum-geometric correlated metals. These contributions are also distinguished by their dependence on the Fermi surface density of states $\rho_F$. The geometric term scales with $\rho_F$ whereas the dispersive term only depends on its derivative \supplement{Supplementary Material, Section V}.

\sectiontitle{Discussion} In summary, we have identified a quantum-geometric contribution to the finite-frequency optical conductivity of clean correlated metals. This effect is generic and will enter as a correction to prior results on optical response in any multiband system \cite{Sharma_DFL, mu2022optical, mu2024adequacy, Mishchenko_Plasmon, Sharma_Plasmon, Gindikin_Intervalley}. However, it is the sole contribution in nearly-parabolic bands such as dilutely-doped metals and is strongly enhanced near a topological band inversion. The key signature of this response in dilute metals is an enhancement of the optical conductivity that leads to a peak when doped to the band inversion scale. 
This implies that the THz optical conductivity of correlated metals can be used as a probe of the quantum geometry of the doping-dependent Fermi surface. Our analysis considers systems with a single Fermi surface in a topologically inverted band. This naturally occurs upon doping materials where the band inversion occurs at one point in the Brillouin zone, such as at the $\Gamma$ point in HgTe quantum wells \cite{Bernevig_2006} or in moir\'e transition metal dichalcogenides \cite{PhysRevLett.122.086402,devakul2021magic,claassen2022ultra,PhysRevLett.132.036501,PhysRevB.109.205121}. Our model also applies to systems with two Fermi surfaces related by time reversal symmetry such as in weakly-doped monolayer transition metal dichalcogenides \cite{PhysRevLett.108.196802}, where the spin index in our model becomes a spin-valley index. Additionally, it would be interesting to extend our analysis to $n$-layer rhombohedral graphene, where the wavefunction winds $n$ times around the $K$ point \cite{Min2008, Zhang2010}. In our model, this  corresponds to the angular momentum of the band inversion. Hence we predict that the low-frequency quantum geometric optical response will drastically increase with $n$.

This work illustrates that quantum geometry can enrich the physics of metals. It would be of interest to interrogate how the structure of Bloch wavefunctions and interactions manifests in other optical responses, particularly in nonlinear and photogalvanic responses and the AC Hall conductivity, where the DC response is known to be a Berry phase contribution \cite{Haldane_Hall}, as well as how it emerges in electromagnetic responses in a semiclassical Boltzmann picture. Additionally, while we treated interactions perturbatively, it would be of great interest to understand how quantum geometry alters the low-frequency electromagnetic response of systems where interactions must be treated non-perturbatively, such as in non-Fermi liquids \cite{lee2018recent,RevModPhys.94.035004}. More broadly, symmetry broken phases in moir\'e materials are preceded by a high-temperature Fermi liquid state. An intriguing question is whether such quantum-geometric probes of the Fermi liquid regime can provide new insights into the role of quantum geometry in the emergent low-temperature symmetry-broken or topologically-ordered phases.

\acknowledgments{
    We thank Jixun Ding, Wai Ting Tai, Spenser Talkington, and Andrew Millis for helpful comments and discussions. D.P.C. and M.C. acknowledge support from the U.S.\ Department of Energy, Office of Basic Energy Sciences, Early Career Award Program, under Award No.\ DE-SC0024494. D.P.C. acknowledges funding from the NSF GRFP under Grant No. DGE-1845298.
}

\sectiontitle{Data availability} The data that support the findings of this study were generated by numerical simulations. The source code used to generate the simulations is publicly available \cite{gFL_Zenodo_Data}.

\bibliography{apssamp}

\appendix

\section{Fermi's Golden Rule Calculation of $\Real \sigma_{\mu\mu}(\omega)$}
In this work we use Fermi's golden rule (FGR) to calculate the optical conductivity as described in \cite{Mishchenko_Plasmon,Maslov_OpticalCond, Gindikin_Intervalley}. This method is based on the observation that the absorption rate of a system when it is coupled to an electric field oscillating at frequency $\omega$ can be related to the dissipative part of the current-current correlation function and hence the real part of the optical conductivity. These observations tell us
\begin{equation*}
\Real \sigma_{\mu, \mu} (\omega) =
 \frac{1}{2 \omega} \Gamma_{\mu}(\omega)
\end{equation*}
where
\begin{align}
    \label{eq:FGR_def}
    \Gamma_{\mu}(\omega) &=
    \frac{2\pi}{4 \Vol^3}
    \sum_{\k, \p, \q} 
    \sum_{\sigma, \sigma'}
    \left|\left[ \boldsymbol{\mathcal{M}}_{\k + \q, \p - \q,
    \p, \k}^{\sigma\sigma'} \right]_\mu \right|^2 \notag\\
    \times~ & \left(1 - e^{-\beta \omega} \right)
    n(\E_\k) n(\E_\p)
    \left (1- n(\E_{\p-\q}) \right)
    \left (1- n(\E_{\k+\q}) \right) \notag\\
    \times~ & ~\delta(\E_{\k + \q} + \E_{\p - \q} - 
    \E_{\k} - \E_{\p} - \hbar \omega ) 
\end{align}
where $\beta$ is the inverse temperature and $\mathcal{M}_{\k + \q, \p - \q, \p, \k}^{\mu; \sigma, \sigma'}$ is the matrix element, to leading order in interactions, that corresponds to processes where two electrons at momenta $\k$ and $\p$ (with spins $\sigma$ and $\sigma'$, respectively) scatter to $\k + \q$ and $\p - \q$. The factor of $1-e^{-\beta \omega}$ originates from a difference of Fermi factors which appear because as electrons are being scattered from $\k, \p \rightarrow \k + \q, \p - \q$ by absorbing a photon of energy $\hbar \omega$, there are also reverse processes $\k + \q, \p -\q \rightarrow \k , \p$ while emitting a photon of energy $\hbar \omega$.

\begin{figure}[h]
\includegraphics[width=\columnwidth]{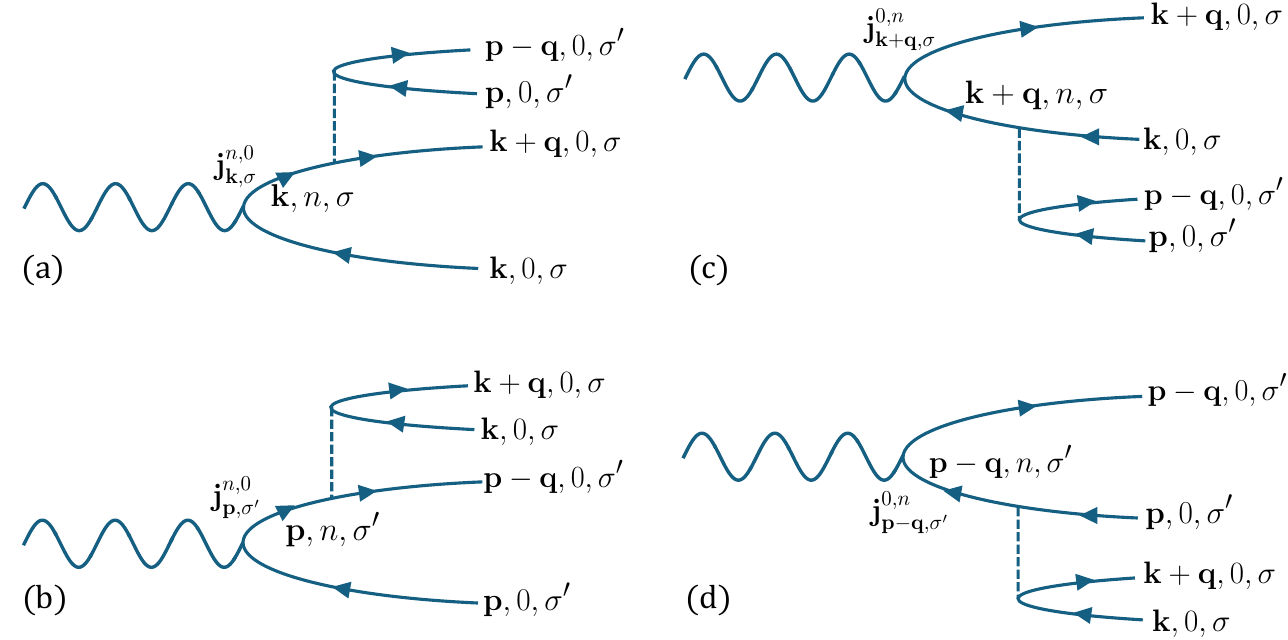}
\caption{\label{fig:SupFigures} Diagrams for the real part of the optical conductivity computed via Fermi's golden rule, with $(\k,n,\sigma)$ denoting momentum, band index, and spin. The virtual intermediate state includes \textit{intra}band ($n=0$) and \textit{inter}band ($n \neq 0$) excitations.}
\end{figure}

The heart of the calculation is to derive $|\mathcal{M}_{\k + \q, \p - \q, \p, \k}^{\mu; \sigma, \sigma'}|^2$ to leading order in the interactions. To do so, it's useful to use a diagrammatic scheme. All permutations of the relevant diagrams are shown in Fig.~\ref{fig:SupFigures}. In contrast to the main text, we do not distinguish diagrams based on whether the intermediate state is an intra- or inter-band excitation. Now, denote $\Delta_{\k}^{n,m} \equiv \E_{n,\k} - \E_{m,\k}$ as the gap between bands $n$ and $m$ at momentum $\k$ (we use $\E_{0,\k} \equiv \E_{\k}$ as short-hand for the active band that hosts the Fermi energy). Using the constraint $\hbar\omega = \E_{\k + \q} + \E_{\p - \q} - \E_{\k} - \E_{\p}$, the resulting matrix elements (labeled following Fig.~\ref{fig:SupFigures}) are quoted as follows:
\begin{align}
\boldsymbol{\mathcal{M}}_a & = \sum_n
\frac{U_{\q}
\langle u_{\k+\q, \sigma}^0 | u_{\k, \sigma}^n \rangle
\langle u_{\p-\q, \sigma'}^0 | u_{\p, \sigma'}^0 \rangle \mathbf{j}^{n, 0}_{\k, \sigma}
}{\E_{0,\k} - \E_{n,\k} + \hbar \omega} 
\nonumber \\
& = U_{\q} \sum_{n}
\langle u_{\k+\q, \sigma}^0 | u_{\k, \sigma}^n 
\rangle
\langle u_{\p-\q, \sigma'}^0 | u_{\p, \sigma'}^0 \rangle \notag\\
& \times \left (
\frac{e \vel_{\k}}{\hbar \omega} \delta_{0,n} + i \frac{e}{\hbar}  \mathbf{A}_{\k, \sigma}^{n,0}  \frac{\Delta_{\k}^{n,0} }{\hbar\omega - \Delta_{\k}^{n,0} } (1-\delta_{0,n})
\right)
\end{align}
\begin{align}
\boldsymbol{\mathcal{M}}_b &  = \sum_n 
\frac{U_{\q}
\langle u_{\k+\q, \sigma}^0 | u_{\k, \sigma}^0 \rangle
\langle u_{\p-\q, \sigma'}^0 | u_{\p, \sigma'}^n \rangle \mathbf{j}^{n, 0}_{\p, \sigma'}
}{\E_{0,\p} - \E_{n,\p} + \hbar \omega} \nonumber \\
&  = U_{\q} \sum_{n}
\langle u_{\k+\q, \sigma}^0 | u_{\k, \sigma}^0 
\rangle
\langle u_{\p-\q, \sigma'}^0 | u_{\p, \sigma'}^n \rangle \notag\\
& \times \left (
\frac{e \vel_{\p}}{ \hbar \omega} \delta_{0,n} + i \frac{e}{\hbar} \mathbf{A}_{\p, \sigma'}^{n,0}  \frac{\Delta_{\p}^{n,0} }{\hbar\omega - \Delta_{\p}^{n,0} } (1-\delta_{0,n})
\right)
\end{align}
\begin{align}
%
\boldsymbol{\mathcal{M}}_c & = \sum_n
\frac{U_{\q}
\langle u_{\k+\q, \sigma}^n | u_{\k, \sigma}^0 \rangle
\langle u_{\p-\q, \sigma'}^0 | u_{\p, \sigma'}^0 \rangle \mathbf{j}^{0, n}_{\k+\q, \sigma}
}{\E_{0,\k+\q} - \E_{n,\k+\q} - \hbar \omega}
\nonumber \\
& = U_{\q} \sum_{n}
\langle u_{\k+\q, \sigma}^n | u_{\k, \sigma}^0 
\rangle
\langle u_{\p-\q, \sigma'}^0 | u_{\p, \sigma'}^0 \rangle  \notag\\
& \times \left (
\frac{e \vel_{\k+\q}}{- \hbar \omega} \delta_{0,n} + 
i \frac{e}{\hbar} \mathbf{A}_{\k+\q, \sigma}^{0,n}  \frac{-\Delta_{\k+\q}^{n,0} }{-\hbar\omega - \Delta_{\k+\q}^{n,0}} (1-\delta_{0,n})
\right)
\end{align}
\begin{align}
%
\boldsymbol{\mathcal{M}}_d & = \sum_n 
\frac{U_{\q}
\langle u_{\k+\q, \sigma}^0 | u_{\k, \sigma}^0 \rangle
\langle u_{\p-\q, \sigma'}^n | u_{\p, \sigma'}^0 \rangle \mathbf{j}^{0, n}_{\p-\q, \sigma'}
}{\E_{0,\p-\q} - \E_{n,\p-\q} - \hbar \omega} \nonumber \\
&  = U_{\q} \sum_{n}
\langle u_{\k+\q, \sigma}^0 | u_{\k, \sigma}^0 
\rangle
\langle u_{\p-\q, \sigma'}^n | u_{\p, \sigma'}^0 \rangle \notag\\
& \times \left (
\frac{e \vel_{\p-\q}}{- \hbar \omega} \delta_{0,n}
+ i \frac{e}{\hbar} \mathbf{A}_{\p-\q, \sigma'}^{0,n}  \frac{-\Delta_{\p-\q}^{n,0}}{-\hbar\omega - \Delta_{\p-\q}^{n,0}} (1-\delta_{0,n})
\right)
\end{align}
where, in order to get the second line of each expression, we have broken up the sum into the interband and intraband contributions.

At low frequencies, $\omega \ll \Delta_{\k}^{n,0}$ is smaller than the energy to generate an interband transition from the Fermi level. Crucially however, an \textit{adiabatic} contribution remains $\frac{\Delta_{\k}^{n,0} }{\hbar\omega - \Delta_{\k}^{n,0} } \approx -1$ and is independent of the intermediate-state detuning. One finds that the total matrix element therefore contains \textit{two} contributions: A kinetic (velocity) contribution
\begin{align}
    \boldsymbol{\mathcal{M}}_{\mathbf{k}, \mathbf{p}, \mathbf{q}}^{\sigma\sigma'; \textrm{vel}}
=
&- \frac{U_\mathbf{q}}{\hbar \omega} 
\left (
\mathbf{v}_{\mathbf{k}+\mathbf{q}}
+ \mathbf{v}_{\mathbf{p}-\mathbf{q}}
- \mathbf{v}_{\mathbf{k}}
- \mathbf{v}_{\mathbf{p}}
\right ) \notag\\
&\times \langle u_{\mathbf{k}+\mathbf{q}, \sigma} | u_{\mathbf{k}, \sigma} \rangle 
\langle u_{\mathbf{p}-\mathbf{q},\sigma'} | u_{\mathbf{p}, \sigma'} \rangle
\end{align}
and a \textit{quantum-geometric} contribution
\begin{align}
    \boldsymbol{\mathcal{M}}_{\mathbf{k}, \mathbf{p}, \mathbf{q}}^{\sigma\sigma'; \textrm{geo}} = i\frac{U_\q}{\hbar} \sum_{n \neq 0} \left\{ \vphantom{\frac{1}{2}} \right. &\mathbf{A}_{\k+\q,\sigma}^{0,n} \braket{u_{\k+\q, \sigma}^n}{u_{\k, \sigma}^0} \braket{u_{\p-\q, \sigma'}^0}{u_{\p, \sigma'}^0} \notag\\
        +~ &\braket{u_{\k+\q, \sigma}^0}{u_{\k, \sigma}^0} \mathbf{A}_{\p-\q,\sigma'}^{0,n}  \braket{u_{\p-\q, \sigma'}^n}{u_{\p, \sigma'}^0} \notag\\
        -~ &\braket{u_{\k+\q, \sigma}^0}{u_{\k, \sigma}^0} \braket{u_{\p-\q, \sigma'}^0}{u_{\p, \sigma'}^n} \mathbf{A}_{\p,\sigma'}^{n,0}   \notag\\
        -~ &\braket{u_{\k+\q, \sigma}^0}{u_{\k, \sigma}^n} \mathbf{A}_{\k,\sigma}^{n,0}  \braket{u_{\p-\q, \sigma'}^0}{u_{\p, \sigma'}^0} \left. \vphantom{\frac{1}{2}} \right\}  \label{eq:MgeoSummedOverBands}
\end{align}
As expected for a quantum-geometric contribution, $\boldsymbol{\mathcal{M}}_{\mathbf{k}, \mathbf{p}, \mathbf{q}}^{\sigma\sigma'; \textrm{geo}}$ can be expressed only in terms of the partially occupied band $0$ that hosts the Fermi level. To do so, we note that:
\begin{align*}
-i \sum_{n\neq0}
\langle u_{\k',\sigma}^0| u_{\k, \sigma}^n  \rangle \mathbf{A}_{\k,\sigma}^{n,0}
& =
\sum_{n\neq0}
\langle u_{\k',\sigma}^0
| u_{\k, \sigma}^n \rangle 
\langle u_{\k, \sigma}^n
|\boldsymbol{\nabla}_{\k} u_{\k, \sigma}^0  \rangle \\
=~&
\langle u_{\k',\sigma}^0| 
\left (1- | u_{\k, \sigma}^0 \rangle 
\langle u_{\k, \sigma}^0 | \right) 
|\boldsymbol{\nabla}_{\k}u_{\k, \sigma}^0  \rangle  \\
=~&
\langle u_{\k',\sigma}^0|
\boldsymbol{\mathcal{D}} u_{\k,\sigma}^0 \rangle
\end{align*}
where we defined a \textit{generalized} covariant derivative $\boldsymbol{\mathcal{D}}$ for the active-band Bloch states:
\begin{align}
    \boldsymbol{\mathcal{D}} \ket{ u_{\k,\sigma}^0 } &= \left( \boldsymbol{\nabla}_{\k} + i \mathbf{A}_{\k,\sigma} \right) \ket{ u_{\k,\sigma}^0 } \\
    \boldsymbol{\mathcal{D}} \bra{ u_{\k,\sigma}^0 } &= \left( \boldsymbol{\nabla}_{\k} - i \mathbf{A}_{\k,\sigma} \right) \bra{ u_{\k,\sigma}^0 }
\end{align}
Here, $\mathbf{A}_{\k,\sigma} \equiv \mathbf{A}_{\k,\sigma}^{0,0}$ is the \textit{intra}band Berry connection. Similarly, one can show that $
i \sum_{n\neq 0} \langle u_{\k,\sigma}^n| u_{\k', \sigma}^0  \rangle \mathbf{A}_{\k,\sigma}^{0,n} = 
\langle \boldsymbol{\mathcal{D}} u_{\k,\sigma}^0|u_{\k',\sigma}^0 \rangle
$. Applying these identities on Eq.~(\ref{eq:MgeoSummedOverBands}), and factoring out $e$, one arrives at the formula for $\boldsymbol{\mathcal{M}}_{\mathbf{k}, \mathbf{p}, \mathbf{q}}^{\sigma\sigma'}$ described in the main text:
\begin{align}
\boldsymbol{\mathcal{M}}_{\mathbf{k}, \mathbf{p}, \mathbf{q}}^{\sigma\sigma'} = &-\frac{1}{\hbar} U_{\q}
\left( \frac{\vel_{\k + \q} + \vel_{\p - \q}
- \vel_{\p} - \vel_\k}{\omega} - \boldsymbol{\mathcal{D}} \right) \notag\\
    &
 ~~~~~\times \langle u_{\k+\q, \sigma}^0 | u_{\k, \sigma}^0 
\rangle
\langle u_{\p-\q, \sigma'}^0 | u_{\p, \sigma'}^0 \rangle 
\end{align}
In addition to these processes, there is an exchange contribution from additional diagrams which differ by a minus sign corresponding to fermion lines crossing. Summing over all diagrams, we arrive at the formula for the optical conductivity presented in the main text, Eq.~(\ref{eq:fullresponse}).

\section{Leading-Order Frequency Scaling of the Optical Conductivity}

In this section, we derive the leading-order frequency scaling of the optical conductivity, presented in the main text. At low frequencies, resonant scattering near the Fermi surface must be decomposed into density, exchange, and pairing channels. The density channel yields a subleading frequency scaling contribution and will be neglected below.

\subsection{Exchange Channel}
The exchange channel describes the scattering of electron pairs $(\k,\sigma)$, $(\p+\dq,\sigma')$ to $(\p,\sigma)$, $(\k+\dq,\sigma')$, where $\dq$ is a small momentum shift away from the Fermi surface. Following a similar procedure to Ref.~\cite{Sharma_DFL}, in order to cleanly expand in terms of $\dq$ it is useful to rewrite the integral expression by introducing an auxiliary energy scale $\Omega$.
\begin{widetext}
\begin{align}
\text{Re} \ \sigma_{\mu\mu}^{\textrm{exc}} \left ( \omega \right) &=
\frac{\pi e^2 (1 - e^{- \beta \omega})}{4\omega} \sum_{\sigma,\sigma'}
\iiint \frac{d\k d\p d\dq}{(2 \pi)^6} \ 
\left |
\left[
\boldsymbol{\mathcal{M}}_{
\k, \p + \dq, \p - \k
}^{\sigma, \sigma'}
-
\delta_{\sigma, \sigma'}
\boldsymbol{\mathcal{M}}_{
\k, \p + \dq, \dq
}^{\sigma, \sigma'}
\right ]_{\mu}
\right |^2 
\nonumber
\\
&
\hspace{1.5cm} \times 
n(\epsilon_{\k})
n(\epsilon_{\mathbf{p+\dq}} )
\left [ 1 - n(\E_{\k+\dq}) \right ]
\left [ 1 - n(\epsilon_{\p}) \right ]
\delta \left (
\E_{\k+\dq} + \E_{\p} - \E_{\p+\dq} -\E_{\k} - \hbar \omega
\right ) 
\\
&=
\frac{e^2 (1 - e^{- \beta \omega})}{256 \pi^5 \omega}
\sum_{\sigma,\sigma'} \iiint d \k d\p d\dq 
\int d \Omega \
\left |
\left [
\boldsymbol{\mathcal{M}}_{
\k, \p + \dq, \p - \k
}^{\sigma, \sigma'}
-
\delta_{\sigma, \sigma'}
\boldsymbol{\mathcal{M}}_{
\k, \p + \dq, \dq
}^{\sigma, \sigma'}
\right ]_{\mu}
\right |^2 
\nonumber
\\
& 
\hspace{1.5cm} \times 
n(\epsilon_{\k})
n(\epsilon_{\mathbf{p}} - \Omega )
\left [ 1 - n(\E_{\k} + \hbar \omega - \Omega) \right ]
\left [ 1 - n(\epsilon_{\p}) \right ]
\nonumber
\\
& \hspace{1.5cm} \times
\delta \left (
\E_{\k+\dq} -\E_{\k} + \Omega - \hbar \omega
\right )
\delta \left (
\E_{\p} - \E_{\p+\dq} - \Omega
\right ).
\end{align}
We now expand in small $\dq$ and perform the integration $\dq$. The delta functions will fix $\dq$ and introduce Jacobian factors. We write $\dq = \delta_\parallel \mathbf{\hat{e}^\p_{\parallel}} + \delta_\perp \mathbf{\hat{e}^\p_{\perp}}$ where $\mathbf{\hat{e}^\p_{\parallel}} = \vel_\p/v_\p$, where $v_\p = |\vel_\p|$ and $\mathbf{\hat{e}^\p_{\perp}} = \mathbf{\hat{z}}\times\mathbf{\hat{e}^\p_{\parallel}}$. With this, the second delta function becomes
\begin{align}
\delta(\E_\p - \E_{\p + \dq} - \Omega)
& \approx
\delta(- \hbar v_p \delta_\parallel - \Omega) = \delta(\hbar v_p \delta_\parallel + \Omega)
\end{align}
This fixes $\delta_\parallel = - \Omega/\hbar v_p$ and contributes a Jacobian factor of $1/\hbar |v_p|$. The first delta function is then 
\begin{align}
\delta(\E_{\k + \dq} - \E_{\k} + \Omega - \hbar \omega)
& \approx
\delta(\hbar \vel_\k \cdot (\delta_\parallel \mathbf{\hat{e}_{\parallel}^\p} + \delta_\perp \mathbf{\hat{e}^\p_{\perp}}) + \Omega - \hbar \omega)
\notag\\
& = \delta(-v_{\k,\parallel}\Omega/v_p + \hbar \delta_\perp v_{\k,\perp} +  \Omega - \hbar \omega) ~.
\end{align}
Integrating over $\delta_{\perp}$ gives a Jacobian factor of $1/\hbar|v_{\k,\perp}|$. As such the overall Jacobian factor is $1/\hbar^2|v_{\k,\perp} v_\p| = 1/\hbar^2\left |\vel_\k \times \vel_\p \right |$. Moreover, the leading low-frequency frequency scaling behavior can be computed by constraining the matrix elements $\boldsymbol{\mathcal{M}}$ to lie on the Fermi surface ($\dq = 0$).
The exchange-channel contribution to the low-frequency optical conductivity therefore becomes
\begin{align}
\Real \sigma_{\mu\mu}^{\textrm{exc}} \left ( \omega \right) \approx
\frac{e^2}{ 256 \pi^5 \omega}
\left ( 
1 - e^{-\beta \omega}
\right ) 
&
\sum_{\sigma,\sigma'} \iint d \k d\mathbf{p} \frac{1}{\hbar^2 
\left |\vel_\k \times \vel_\p \right |
}
\left |
\left (
\boldsymbol{\mathcal{M}}_{
\k, \p , \p - \k
}^{\sigma, \sigma'}
-
\delta_{\sigma, \sigma'}
\boldsymbol{\mathcal{M}}_{
\k, \p , 0
}^{\sigma, \sigma'}
\right )_{\mu}
\right |^2 
\nonumber
\\
&
\times \int d \Omega \
n(\epsilon_{\k})
n(\epsilon_{\mathbf{p}} - \Omega )
\left [ 1 - n(\E_{\k} + \hbar \omega - \Omega) \right ]
\left [ 1 - n(\epsilon_{\p}) \right ]
~.
\end{align}
We now convert this into a Fermi surface integral. For $\k$ close to the Fermi surface, we decompose $d\k$ into the change along and perpendicular to the Fermi surface, $d\k_{FS}$ and $d\k_{\perp}$ respectively. This permits a change of variables $d\k = d \k_{FS} d\k_{\perp} = 
1/\hbar v_\k d \k_{FS} d \E_{\k}$,
and the same for $\p$. We find:
\begin{align}
\Real \sigma_{\mu\mu}^{\textrm{exc}} \left ( \omega \to 0\right) =
&~ \frac{e^2 (1 - e^{-\beta \omega})}{ 256 \pi^5 \hbar^4 \omega} 
\iint\limits_{\textrm{FS}^*} d \k d \p \
\frac{1}{ v_\k v_\p
\left |\vel_\k \times \vel_\p \right |
}
\left |
\left (
\boldsymbol{\mathcal{M}}_{
\k, \p , \p - \k
}^{\sigma, \sigma'}
-
\delta_{\sigma, \sigma'}
\boldsymbol{\mathcal{M}}_{
\k, \p, 0
}^{\sigma, \sigma'}
\right )_{\mu}
\right |^2 
\nonumber
\\
&
\times 
\iiint d \E_\k d \E_\p  d \Omega~
n(\epsilon_{\k})
n(\epsilon_{\mathbf{p}} - \Omega )
\left [ 1 - n(\E_{\k} + \hbar \omega - \Omega) \right ]
\left [ 1 - n(\epsilon_{\p}) \right ] ~.
\end{align}
Crucially, this integral factorizes into a part which is an integral over the Fermi surface ($\textrm{FS}^*$, with a cutoff imposed for momentum points with divergent scattering phase space, as explained below) and an integral over Fermi functions which we can evaluate. At zero temperature, using
\begin{align}
\lim_{\beta \to \infty} \left ( 
1 - e^{-\beta  \omega}
\right )  
\iiint d \E_\k d \E_\p d \Omega~
n(\epsilon_{\k})
n(\epsilon_{\mathbf{p}} - \Omega )
\left [ 1 - n(\E_{\k} + \hbar \omega - \Omega) \right ]
\left [ 1 - n(\epsilon_{\p}) \right ]
= \frac{(\hbar\omega)^3}{6}
\end{align}
we finally obtain the zero-temperature leading order frequency scaling contribution in the exchange channel:
\begin{align}
    \Real \sigma_{\mu\mu}^{\textrm{exc},T=0} \left ( \omega \to 0\right) =
&~ \frac{e^2}{h} ~ \omega^2 ~ \frac{1}{ 48 (2\pi)^4} 
\iint\limits_{\textrm{FS}^*} 
\frac{d \k d \p }{ v_\k v_\p
\left |\vel_\k \times \vel_\p \right |
}
\left |
\left (
\boldsymbol{\mathcal{M}}_{
\k, \p , \p - \k
}^{\sigma, \sigma'}
-
\delta_{\sigma, \sigma'}
\boldsymbol{\mathcal{M}}_{
\k, \p , 0
}^{\sigma, \sigma'}
\right )_{\mu}
\right |^2 
\end{align}
The denominator $|\vel_\k \times \vel_\p |$ effectively measures the phase space for scattering processes and apparently diverges for isolated momentum pairs $(\k, \p)$ on the Fermi surface, e.g. $(\k, -\k)$ for circular Fermi surfaces. This divergence is in principle cured via a higher-order expansion in $\dq$ that encodes the curvature of the Fermi surface and can lead to a logarithmic $\log \hbar\omega/\E_F$ scaling correction to the optical conductivity. To simplify this computation, we instead impose a frequency-dependent cutoff around these points to regularize the integral, and denote such regularized Fermi surface integrals as covering a domain $\textrm{FS}^*$. We discuss the explicit computation for circular Fermi surfaces below.

\subsection{Pairing Channel}

We now repeat this process in the pairing channel, where a pair of electrons $(\k,\sigma)$, $(-\k+\dq,\sigma')$ scatters to $(\p,\sigma)$, $(-\p+\dq,\sigma')$ with small momenta $\dq$:
\begin{align}
\Real \sigma_{\mu\mu}^{\textrm{pair}} \left ( \omega \right) =
\frac{\pi e^2 (1 - e^{-\beta \omega})}{4\omega} 
& \sum_{\sigma,\sigma'} \iiint \frac{d \k d \mathbf{p} d\dq}{(2\pi)^6} \ 
\left |
\left (
\boldsymbol{\mathcal{M}}_{
\k, -\k +\dq, \p - \k
}^{\sigma, \sigma'}
-
\delta_{\sigma, \sigma'}
\boldsymbol{\mathcal{M}}_{
\k, -\k +\dq, \dq -\p - \k
}^{\sigma, \sigma'
}
\right )_{\mu}
\right |^2 
\nonumber
\\
&
\times 
n(\epsilon_{\k})
n(\epsilon_{-k + \dq} )
\left [ 1 - n(\E_{-\p+\dq}) \right ]
\left [ 1 - n(\epsilon_{\p}) \right ]
\nonumber
\\
& \times
\delta \left (
\E_{\p} + \E_{-\p+\dq} - \E_{-\k+\dq} -\E_{\k} - \hbar \omega
\right ) 
\nonumber
\\
=
\frac{e^2 (1 - e^{-\beta \omega})}{256\pi^5 \omega}
& \iiint d \k d \mathbf{p} d\dq 
\int d \Omega \
\left |
\left (
\boldsymbol{\mathcal{M}}_{
\k, -\k +\dq, \p - \k
}^{\sigma, \sigma'}
-
\delta_{\sigma, \sigma'}
\boldsymbol{\mathcal{M}}_{
\k, -\k +\dq, \dq -\p - \k
}^{\sigma, \sigma'
}
\right )_{\mu}
\right |^2 
\nonumber
\\
&
\times 
n(\epsilon_{\k})
n(\Omega - \hbar \omega - \E_{\k})
\left [ 1 - n(\Omega - \E_\p) \right ]
\left [ 1 - n(\epsilon_{\p}) \right ]
\nonumber
\\
& \times
\delta \left (
\Omega - \hbar \omega - \E_{\k - \dq} - \E_\k
\right )
\delta \left (
\E_{\p} + \E_{\p-\dq} - \Omega
\right ) . 
\end{align}
where we used $\E_{\k} = \E_{-\k}$ in a centrosymmetric metal. For small-momentum $\dq$ scattering at low frequencies, the second delta function can be approximated as
\begin{align}
\delta(\E_\p + \E_{\p - \dq} - \Omega) & \approx
\delta\left( 2 \E_\p - \hbar \vel_\p \cdot \dq - \Omega \right)
=
\delta(2 \E_\p - \hbar v_\p  \delta_\parallel - \Omega)
\end{align}
This fixes $\delta_\parallel = (2 \E_\p - \Omega )/\hbar v_\p $ upon integration.
The first delta function argument becomes
\begin{align}
\delta(\Omega - \hbar \omega - \E_{\k - \dq} - \E_{\k}) & \approx
\delta(\Omega - \hbar \omega - 2 \E_\k + \hbar \vel_\k \cdot \dq )
\\ 
& = 
\delta\left(\Omega - \hbar \omega - 2 \E_\k + \hbar \vel_\k \cdot (\delta_\parallel \mathbf{\hat{e}_{\parallel}^\p} + \delta_\perp \mathbf{\hat{e}^\p_{\perp}}) \right)
\\
& = 
\delta\left( \Omega - \hbar \omega - 2 \E_\k +  (2 \E_\p - \Omega) v_{\k, \parallel} / v_\p + \hbar \delta_\perp  v_{\k, \perp} \right) ~.
\end{align}
Integrating again yields a Jacobian factor $1/\hbar^2 |v_{\k,\perp} v_\p|$. We then repeat the low-frequency-limit approximation that the matrix element must lie on the Fermi surface (that is we evaluate it for $\dq = 0$) and perform the integrals over frequency. At zero temperature, the pairing-channel contribution to the low-frequency optical conductivity finally becomes
\begin{align}
\Real \sigma_{\mu\mu}^{\textrm{pair},T=0} \left ( \omega \right) = \frac{e^2}{h} ~ \omega^2 ~ 
\frac{1}{48 (2\pi)^4} 
& \sum_{\sigma,\sigma'} \iint\limits_{\textrm{FS}^*} 
\frac{d \k d \p}{ v_\k v_\p
\left |\vel_\k \times \vel_\p \right |
}
\left |
\left (
\boldsymbol{\mathcal{M}}_{
\k, -\k , \p - \k
}^{\sigma, \sigma'}
-
\delta_{\sigma, \sigma'}
\boldsymbol{\mathcal{M}}_{
\k, -\k, -\p - \k
}^{\sigma, \sigma'
}
\right )_{\mu}
\right |^2 
\end{align}

\subsection{Leading-Order Frequency Scaling at Zero Temperature}

Combining the exchange and pairing channel contributions, we find that the real part of the low-frequency optical conductivity has a succinct expression as a Fermi surface property:
\begin{align}
    \Real \sigma_{\mu\mu}(\omega) = \frac{e^2}{h} \left( \alpha_\mu^{\textrm{exc}} + \alpha_\mu^{\textrm{pair}} \right) ~ \omega^2 
\end{align}
where
\begin{align}
    \alpha_{\mu}^{\textrm{exc}} &= \frac{1}{48 (2\pi)^4} 
 \sum_{\sigma,\sigma'} \iint\limits_{\textrm{FS}^*} 
\frac{d \k d \p}{ v_\k v_\p \left |\vel_\k \times \vel_\p \right |} 
\left |
\left[
\boldsymbol{\mathcal{M}}_{
\k, \p , \p - \k
}^{\sigma, \sigma'}
-
\delta_{\sigma, \sigma'}
\boldsymbol{\mathcal{M}}_{
\k, \p, 0
}^{\sigma, \sigma'}
\right]_{\mu}
\right |^2 
\\
\alpha_{\mu}^{\textrm{pair}} &= \frac{1}{48 (2\pi)^4} 
 \sum_{\sigma,\sigma'} \iint\limits_{\textrm{FS}^*} 
\frac{d \k d \p}{ v_\k v_\p \left |\vel_\k \times \vel_\p \right |} 
\left |
\left [
\boldsymbol{\mathcal{M}}_{
\k, -\k , \p - \k
}^{\sigma, \sigma'}
-
\delta_{\sigma, \sigma'}
\boldsymbol{\mathcal{M}}_{
\k, -\k, -\p - \k
}^{\sigma, \sigma'
}
\right ]_{\mu}
\right |^2 
\end{align}

\subsection{Logarithmic Scaling for Circular Fermi Surfaces}

For a circular Fermi surface we can 
write Eq.~(\ref{eq:sigma_FS}) of the main text in polar coordinates and impose the cut off as follows:
\begin{align}
\Real \sigma_{\mu\mu}(\omega) =
\frac{e^2}{h^3} \omega^2 ~\frac{1}{48 (2\pi)^2} 
~ \frac{k_F^2}{v_F^4}
\int_{-\pi}^\pi d\theta
\int_{-\pi + \omega/\Lambda}
^{\pi - \omega/\Lambda}
d\phi
\frac{~ U^2_{\k-\p} }{ 
\left | \sin\phi \right | } ~\mathcal{G}_{\k,\p}^{\mu}  
\end{align}
where $v_F$ is the Fermi velocity, $\phi$ is the angle between $\k$ and $\p$, and $\theta$ is the angle $\k$ makes with respect to the $k_x$ axis. Near the cut off, the integrand diverges as $1/(\pi - \phi)$. As such, this divergence  gives rise to a term of the form $\tilde\beta \omega^2 \log \omega/\Lambda$ after performing the integrals. Away from the divergence the scattering processes will give contributions which scale as $\omega^2$. 

Overall, the implementation of the cut-off in the integral results in the optical conductivity taking the form:
\begin{align*}
\Real \sigma(\omega) = \tilde\alpha \omega^2 - \tilde\beta \omega^2 \log \omega/\Lambda
\end{align*}
where $\tilde\alpha = \tilde\alpha(\omega/\Lambda)$ formally depends on the cutoff due to the bounds of integration. Since we are only considering the conductivity to leading order in $\omega$ we are only concerned with 
$\tilde\alpha(\omega/\Lambda = 0)$. To convert this into the scaling form used in the main text we rearrange this as
\begin{align*}
\lim_{\omega \rightarrow 0} \Real \sigma(\omega)
& = \tilde\alpha (0)\omega^2 - \tilde\beta \omega^2 \log \omega/\Lambda\\
&= \tilde\alpha (0)\omega^2 - \tilde\beta \omega^2
(\log \hbar\omega/\E_F - \log \hbar\Lambda/\E_F) \\
& = \left(
\tilde\alpha(0)
+ \tilde\beta \log \hbar\Lambda/\E_F
\right ) \omega^2 - \tilde\beta \omega^2
\log \hbar\omega/\E_F
\end{align*}
which is the equivalent to the form stated in the main text upon identifying $\alpha = \tilde\alpha(0)
+ \tilde\beta \log \Lambda/\E_F$ and $\beta = \tilde\beta$.

For the model considered in the main text [Eq.~(\ref{eq:chiral-inversion})] in the parabolic limit we find that
\begin{align*}
\beta = 
\alpha_0
\frac{\pi^2}{3}
\frac{
\left ( 1 + (-1)^{m+1} \right )
m^2 \lambda^{4 m }
(\lambda^{2m} - 1 )^2
}
{(1+ \lambda^{2m})^6 }
\end{align*}
where $\alpha_0 = \frac{1}{(2 \pi)^6} \hbar e^2 U^2 k_F^4/\E_F^4$.

\end{widetext}

\section{Monte Carlo Evaluation of Optical Conductivity}
To study the competition between quantum-geometric and kinetic contributions away from Galilean invariance, we evaluated Eq.~(\ref{eq:fullresponse}) of the main text for the higher angular momentum band inversion model  using Monte Carlo integration. For circular Fermi surfaces, $\k, \p,$ and $ \q$ can be parameterized 
by their lengths and their polar angles. The integration over $|\q|$ is fixed via the delta function and performed analytically.  Here, the $\propto k^2 + k^4$ in the dispersion presented in the main text enables an analytical solution for the zeros of the delta function, as well as the Jacobian factor that is contributed from integrating over $|\q|$. The remaining lengths and angles are sampled during Monte Carlo integration.

To determine the frequency scaling behavior, we compute the optical conductivity at frequencies ranging from $\hbar \omega/\E_F = 10^{-3}$ to $\hbar \omega/\E_F = 10^{-1}$ where $\E_F$ is the Fermi energy, using $10^8$ samples at each frequency to ensure the integral is converged. From this, we fit the resulting optical conductivities to the scaling form $\alpha \omega^2 - \beta \omega^2 \log(\hbar\omega/\E_F)$. The results are presented in Fig.~\ref{fig:geocon} and Fig.~\ref{fig:fullCond} of the main text.

\section{Analytical Properties of $\mathcal{G}$}

To build more physical intuition behind the optical response we present explicit forms of $\mathcal{G}$ for various scattering processes. First we consider scattering processes with small momentum transfer. For such processes $\mathcal{G}$ can be expanded in powers of the momentum transfer, $\p - \k$, which yields
\begin{align*}
\mathcal{G}^\mu_{\k, \p} = 
8 \sum_\sigma
\left (p^\nu - k^\nu \right )
\left (p^\rho - k^\rho \right )
\mathcal{F}^{\mu \nu}_{\k \sigma}
\mathcal{F}^{\mu \rho}_{\k \sigma} +
\mathcal{O} \left ( \left ( \p - \k \right)^3 \right ) 
\end{align*}
where $\mathcal{F}^{\mu \nu}_{\k, \sigma} = \partial^\mu A^\nu_{\k, \sigma} - \partial^\nu A^\mu_{\k, \sigma} $  is the Berry curvature, which is proportional to the imaginary part of the quantum geometric tensor. This term appears due to the spinful exchange and pairing channels whereas the spinless exchange channel leads to a contribution at higher order in the momentum transfer. For a spinless, or spin polarized, Fermi surface the contribution is instead

\begin{align*}
\mathcal{G}^\mu_{\k, \p}
= 
\partial_{k^\mu} g^{\nu \rho}_{\k}
\partial_{k^\mu} g^{\sigma \tau}_{\k}
q^\nu q^\rho q^\sigma q^\tau 
+ \mathcal{O}\left ( q^5
\right )
\end{align*}
where $g^{\nu \rho}$ is the quantum metric and we have defined $ \q = \p - \k$ in order to simplify the expression.

Now we consider the case of scattering from $\k \rightarrow -\k$ which contributes to the logarithmic scaling of the optical conductivity. For these scattering processes we have that 
\begin{align*}
\mathcal{G}^\mu_{\k, -\k} = 4 \sum_\sigma
& \left | 
\braket{u_{-\k \sigma}}{u_{\k \sigma}}
\right |^4 \\
& \times \left |
2  A^\mu_{\k \sigma} + \left (
\partial_{\k^\mu} + \partial_{-\k^\mu} 
\right)
\textrm{Arg} \braket{
u_{-\k \sigma}
}{
u_{\k \sigma}
}
\right|^2
\end{align*}
where we have used time reversal and inversion symmetry to simplify the expression. This form demonstrates that the $2 k_F$ processes depend on the overlap between the Bloch wavefunctions at opposite momentum as well as another gauge invariant quantity that is the sum of the Berry connection alongside a term which cancels off the gauge dependent part of the Berry connection.

Interestingly, we can generalize the second factor to any scattering between two momenta $\k$ and $\p$, not just $2 k_F$ scattering. This takes the form
$
S^\mu_{\k,\p,\sigma} = A^\mu_{\k \sigma} - A^\mu_{\p \sigma} + \left (
\partial_{\k^\mu} + \partial_{\p^\mu} 
\right)
\textrm{Arg} \braket{
u_{\p \sigma}
}{
u_{\k \sigma}
}.$ Remarkably, this generalized form is also gauge invariant. For gauge transformations of the form
$\ket{u_{\k \sigma}} \rightarrow e^{i \phi_\k} \ket{u_{\k \sigma}}  $ the Berry connections transform as $A^\mu_{\k \sigma} \rightarrow A^\mu_{\k \sigma} - \partial_{\k^\mu}\phi_{\k \sigma}$ and the argument transforms as 
$\textrm{Arg} \braket{ u_{\p \sigma} }{ u_{\k \sigma}} \rightarrow \textrm{Arg} \braket{ u_{\p \sigma} }{ u_{\k \sigma}} - \phi_{\p, \sigma} + \phi_{\k, \sigma}.
$ From this we can see that the whole term transforms as
\begin{align*}
S^\mu_{\k,\p,\sigma} & \rightarrow 
S^\mu_{\k,\p,\sigma} - 
\partial_{\k^\mu}\phi_{\k \sigma} 
+ \partial_{\p^\mu}\phi_{\p \sigma} \\
& \mspace{50 mu}
+ \left ( \partial_{\k^\mu} + \partial_{\p^\mu} \right ) \left ( 
\phi_{\k, \sigma}  -  \phi_{\p, \sigma} \right ) \\
& = S^\mu_{\k,\p,\sigma}
\end{align*}

We make the observation that this quantity has the same form as the shift vector but depends on the Bloch wavefunctions at different momentum points, as opposed to different bands. Interestingly, this matrix element is very similar to the expression for the shift in the center of mass of a wave packet when scattered by an impurity from momentum $\k$ to $\p$ as discussed in \cite{Sinitsyn2006}. This suggests that the quantum geometric quantity $\mathcal{G}_{\k,\p}$, at all $\k$ and $\p$, can be understood as a real-space shift in the center of mass of a pair of electrons when scattered by interactions. We defer the discussion of such a real-space formulation to future investigations as it is beyond the scope of this work.

\section{
Distinguishing Geometric and Dispersive Contributions}

In the main text, we showed that the contributions from quantum geometry and the band dispersion can be distinguished via their dependence on doping. In particular, a key signature of the quantum geometric contribution, shown in Fig. 4, is an observable maximum in optical absorption as a function of doping, whereas conventional dispersive contributions exhibit a slow and monotonic onset when doping away from the band bottom. In this section, we show that both processes depend differently on the density of states $\rho(k_F)$ at the Fermi surface.

For an isotropic two-dimensional dispersion, the density of states per spin at momentum $k$ is given by, $\rho(k) = \frac{1}{2 \pi \hbar} \frac{k_F}{v_F}$. From this, we see that the quantum geometric optical conductivity scales as
\begin{align*}
\sigma_{geo}(\omega) & = \frac{e^2}{h^3} \omega^2 ~\frac{1}{48 (2\pi)^2} \oint\limits_{\textrm{FS}^*} \frac{d\k d\p ~ U^2_{\k-\p} }{ v_\k v_\p
\left |\vel_\k \times \vel_\p \right | } ~\mathcal{G}_{\k,\p}^{\mu} \\
& \sim
\frac{e^2}{h^3} ~\frac{k_F^2}{v_F^4} U^2 \omega^2 \oint\limits_{\textrm{FS}^*} \frac{d\phi d\theta 
}{ \sin \phi } ~\mathcal{G}_{\k,\p}^{\mu} \\
& \sim
\frac{e^2}{h} ~\frac{\rho(k_F)^2 \ell_g^2}{v_F^2} U^2 \omega^2 
\end{align*} 
\noindent where, $\ell_g$ is the length scale that comes from evaluating the integral of $\mathcal{G}$ over the Fermi surface. It can be thought of an emergent length scale which originates from how the Bloch wavefunctions change at the Fermi surface. For simplicity we have assumed the interaction is highly screened $U_q = U$. Finite screening lengths or long-ranged interactions will change the effective values of $U$ but only weakly modify $\ell_g$ (insets of Fig. 3 of the main text), and will not alter how the conductivity scales with the density of states at the Fermi surface.

To derive the scaling of the dispersive term, we follow \cite{Sharma_DFL} and approximate $\vel_{\k + \q} + \vel_{\p - \q} - \vel_\p - \vel_\k \sim w \omega^2/k_F$. Here $w = k_F^2 f'(k_F)/\hbar v_F$ where $f(\k) = |\vel|/|\k| \sim 1/\hbar \rho(k)$. As such we can repeat the analysis above to find that

\begin{align*}
\sigma_{disp}\left(\omega \right) \sim 
\frac{e^2}{h} \frac{ 1}{v_F^2}  \left( \rho'(k_F) \right )^2 U^2 \omega^2.
\end{align*} 
\noindent Hence, the dispersive and geometric contributions can be distinguished in an actual material by observing how the optical response depends on the density of states at the Fermi surface.

\end{document}